\documentclass[12pt]{article}

\usepackage{makecell}
\usepackage{amsmath}
\usepackage{amssymb}
\usepackage{slashed}
\usepackage{amsfonts}
\usepackage{graphicx}
\usepackage[dvipsnames]{xcolor}
\usepackage[colorlinks=true,urlcolor=blue,linkcolor=blue,citecolor=blue]{hyperref}
\usepackage{cite}
\usepackage{cancel}
\usepackage{booktabs}
\usepackage{caption}
\usepackage{subcaption}
\usepackage{float}
\usepackage{soul}

\usepackage[top=1.135in,bottom=1.17in,left=1.16in,right=1.16in]{geometry}
\numberwithin{equation}{section}

\allowdisplaybreaks

\newcommand{\be}{\begin{equation}}
\newcommand{\ee}{\end{equation}}

\usepackage{graphicx} 
\usepackage{amssymb,amsthm,amsmath,amsfonts}
\usepackage{makecell}
\usepackage{bm}
\usepackage{multicol}
\usepackage{appendix}
\usepackage{orcidlink}

\usepackage{cite}
\usepackage[T1]{fontenc}

\begin{document}
\thispagestyle{empty}

\begin{center}

{\bf\Large \boldmath HEFT approach to investigate the muon $g$-2 anomaly at a muon collider}

\vspace{50pt}

Fabiola Fortuna~\orcidlink{0000-0002-8938-7613}$^{1}$, Juan Manuel Márquez~\orcidlink{0009-0007-3354-2497} $^{2}$ and Pablo Roig~\orcidlink{0000-0002-6612-7157} $^{2,3}$

\vspace{16pt}
{$^{1}$\it Instituto de F\'isica, Universidad Nacional Aut\'onoma de M\'exico, AP 20-364, Ciudad de M\'exico 01000, M\'exico}\\
{$^{2}$\it Departamento de F\'isica, Centro de Investigaci\'on y de Estudios Avanzados del Instituto Polit\'ecnico Nacional} \\
{\it Apartado Postal 14-740, 07000 Ciudad de México, M\'exico}\\
{$^{3}$\it IFIC, Universitat de Val\`encia – CSIC, Catedr\'atico Jos\'e Beltr\'an 2, E-46980 Paterna, Spain} \\
\vspace{16pt}

\vspace{30pt}

{\tt}

\vspace{20pt}

\end{center}

\begin{abstract}
In a previous work \cite{Buttazzo:2020ibd}, it has been pointed out that heavy new physics (NP) contributions to lepton dipole moments and high-energy cross-sections of lepton pairs into Higgs and $\gamma/Z$ bosons are connected, if the electroweak symmetry breaking is realized linearly. As a consequence, a multi-TeV muon collider would provide a unique test of NP in the muon $g$-2 through the study of high-energy processes such as $\mu^+ \mu^-\to h +\gamma /Z$.
Since the analysis involved a SMEFT approach to Higgs-processes, it could also be studied by the more general HEFT formulation. We compute the modification of the high-energy cross-section $\mu^+ \mu^-\to h +\gamma/Z$ and $h\to \mu^{+} \mu^{-} +\gamma /Z$ decay using the dimension six HEFT Lagrangian and compare them with the SMEFT analysis. We find that, within the current HEFT analysis, there are plausible scenarios where the HEFT approach could lead to a higher sensitivity to test the NP contributions in
the muon $g$-2. However, a more precise knowledge of the new HEFT parameters is needed for a definite conclusion, which motivates the search for complementary measurements.
\end{abstract}

\newpage


\section{Introduction}

The formalism of Effective Field Theories (EFT) represents an optimal tool for studying the phenomenology of the Standard Model (SM) and beyond, since it provides a comprehensive and easy description of nature at certain physical scales. 

In particular, the Standard Model Effective Field Theory (SMEFT) \cite{Buchmuller:1985jz,Grzadkowski:2010es} provides a very powerful framework, maintaining invariance under Lorentz and SM gauge symmetries, while adhering to an expansion in canonical mass dimensions. This allows for a comprehensive and model-independent investigation of potential physics beyond the SM. By incorporating higher-dimension operators, SMEFT facilitates the exploration of new physics that might arise from the interactions of massive particles as yet undetected.

While the SMEFT formalism treats the Higgs as an elementary $SU(2)$ doublet, there are scenarios where the Higgs does not strictly belong to this representation which are still allowed within the current experimental accuracy, such as the Composite Higgs models \cite{Kaplan:1983fs,Kaplan:1983sm,Banks:1984gj,Agashe:2004rs,Gripaios:2009pe}. This makes intriguing to identify observables that can distinguish between these different possibilities.

The scenarios where the Higgs does not belong to an $SU(2)_L$ doublet are described by the so-called Higgs EFT (HEFT) \cite{Appelquist:1980vg,Appelquist:1980ae,Longhitano:1980iz,Longhitano:1980tm,Feruglio:1992wf,Grinstein:2007iv,Alonso:2012px,Buchalla:2013rka,Alonso:2015fsp,Alonso:2016oah}, where the physical Higgs $h$ is treated as a singlet under the SM gauge group, separately from the three electroweak Goldstone bosons, which are collectively described by the Higgs doublet in SMEFT. Consequently, the number of independent operators at any given order in the expansion is significantly higher than in SMEFT. Additionally, in the HEFT framework, the dependence on the $h$ field is customarily encoded in generic functions, $\mathcal{F}_i(h)$, which serve as the building blocks for the construction of the effective operators.

The differences between the SMEFT and HEFT approaches can lead to distinct phenomenological outcomes when analyzing processes involving the Higgs sector. Numerous discussions, such as those in Refs. \cite{Brivio:2013pma,Gavela:2014vra}, have identified signatures that may help distinguish between these two frameworks. Examples include studies in Dark Matter \cite{Brivio:2015kia} and the scattering of the longitudinal components of gauge bosons \cite{Murayama:2014yja,Delgado:2013hxa}. For a comprehensive review of both formalisms, we recommend Refs.\cite{Brivio_2016,Isidori:2023pyp}.

With this motivation, different classes of EFT’s have been widely used to describe all kinds of NP contributions to a variety of different observables. Specially interesting is the case of the anomalous magnetic moment of the muon, since it has been an enduring hint for NP (see, however, the recent Ref.~\cite{Boccaletti:2024guq}). The latest measurements by the Muon $g$-2 collaboration at Fermilab \cite{Muong-2:2023cdq,Muong-2:2021ojo}, when combined with earlier results from the Brookhaven E821 experiment \cite{Muong-2:2006rrc}, indicate a deviation of 5.1$\sigma$ from the SM prediction \cite{Aoyama:2020ynm}:
\begin{equation} \label{eq:delta_a_mu}
\Delta a_\mu= a_\mu^{\text{exp}}-a_\mu^{\text{SM}}=
249(48)\times 10^{-11}
.
\end{equation}
However, due to the high level of precision required, it is extremely challenging to conduct an independent test of NP in the muon $g$-2 that is not influenced by (mostly) hadronic and experimental uncertainties. Furthermore, there is a discrepancy in the theoretical predictions of the $g$-2 muon anomaly, depending on the data used as inputs in the calculation. The deviation of $5.1\sigma$ mentioned before, was obtained using $e^{+}e^{-}$ data-driven methods (which take experimental hadron cross-section data as input), but there is also the alternative of using $\tau$ data-driven \cite{Miranda:2020wdg,Masjuan:2024mlg,Davier:2023fpl,Castro:2024prg} and lattice QCD computations \cite{Borsanyi:2020mff}, which in fact reduces the tension between the theoretical and experimental values to 2.0 $\sigma$ and 1.5 $\sigma$, respectively (less than one $\sigma$ in \cite{Boccaletti:2024guq}). The latest CMD-3 measurement of $\sigma(e^+e^-\to\pi^+\pi^-)$ \cite{CMD-3:2023alj,CMD-3:2023rfe} also points in this direction.

Recently, it was emphasized that heavy NP contributions in lepton dipole moments and high-energy cross-sections of lepton pairs into Higgs bosons and $\gamma/Z$ are connected, under the assumption that the electroweak symmetry is realized linearly. Ref.~\cite{Buttazzo:2020ibd} exploits this property, and finds that a multi-TeV muon collider \cite{Delahaye:2019omf,MICE:2019jkl,Bartosik:2020xwr,Capdevilla:2020qel,Yin:2020afe,AlAli:2021let,Huang:2021nkl,MuonCollider:2022xlm,Black:2022cth,Accettura:2023ked,Arakawa:2022mkr} would enable us to directly probe NP contributions to the muon $g$-2 \cite{Athron:2021iuf}, not hampered by the hadronic uncertainties that affect the SM prediction. This is because measuring the high-energy cross section for the processes $\mu^+\mu^-\to h+\gamma/Z$ would give equivalent information to measuring $\Delta a_\mu$, within the EFT framework.

More importantly, the cross section for muon scattering processes increases with the collider energy, as these are driven by effective operators. Then a high-energy measurement with $\mathcal{O}(1)$ precision will be sufficient to disentangle NP effects from the SM background. This approach offers sensitivity to the muon's magnetic moment that is several orders of magnitude beyond other projected collider constraints, as discussed in \cite{Buttazzo:2020ibd}, where this interplay of the high-energy and high-intensity frontiers of particle physics was explored in detail within the SMEFT formalism. 

As mentioned above, the required high-energy processes involve Higgs boson interactions, as well as rare decays of the Higgs. It is therefore interesting to consider how the analysis would differ if conducted within the HEFT formalism, where the non-linear realization of the symmetry could lead to different results, potentially enhancing the sensitivity to $\Delta a_\mu$ in a future muon-collider.

This work is structured as follows: after a review of the SMEFT approach in section \ref{sec:2}, we recall the dimension-six HEFT Lagrangian that describes the processes under consideration in section \ref{sec:3}. Then, in sections \ref{sec:4} and \ref{sec:5}, we discuss the HEFT computation for the $\mu^+\mu^-\to h\gamma$ and $\mu^+\mu^-\to hZ$ processes respectively, and analyze the main differences between SMEFT and HEFT approaches, considering the $\Delta a_\mu$ sensitivity that a muon collider could have, using different sets of well-motivated values of the HEFT Wilson coefficients. Our conclusions are given in section \ref{sec:7}. After that, in appendix \ref{App:HDecay}, we analyze the related Higgs rare decays as an alternative tool to study new physics effects from the HEFT approach, and discuss the corresponding results.

\section{SMEFT approach}
\label{sec:2}

In this section we give a brief overview on how to investigate the muon $g$-2 anomaly at a muon collider within the SMEFT framework (see e.g. refs.~\cite{Arzt:1992wz,Einhorn:2001mf,Fajfer:2021cxa,Aebischer:2021uvt,Crivellin:2021rbq,Cirigliano:2021peb,Dermisek:2023nhe} for earlier EFT analysis). Then, in the next sections, we will point out the main differences between SMEFT and HEFT approaches when addressing this topic. 
Before beginning with the HEFT computation, we believe it is useful to highlight the main properties and ideas of Ref.~\cite{Buttazzo:2020ibd}, whose work serves as motivation for this analysis. We strongly recommend the reader to take a look to their work for full and explicit details on the summary that we present here. 

As commented in Ref.~\cite{Buttazzo:2020ibd}, the observed muon discrepancy can be accommodated by NP effects that arise from the dimension-6 dipole operators\footnote{We focus here only on the operators where the Higgs boson participates. $\ell$ is the lepton weak doublet and $e$ the corresponding singlet, both for the three SM families. $B_{\mu\nu}$ is the $U(1)$ field-strength tensor and $W_{\mu\nu}^I$ the corresponding one for $SU(2)$, with $I=1,2,3$ and $2\tau^I=\sigma^I$ the Pauli matrices.} 
\begin{equation} \label{eq:lag-smeft1}
    \mathcal{L}=\frac{C^\ell_{eB}}{\Lambda^2} \left(\bar{\ell}_L \sigma^{\mu\nu} e_R\right) H B_{\mu\nu} + \frac{C^\ell_{eW}}{\Lambda^2} \left(\bar{\ell}_L \sigma^{\mu\nu} e_R\right) \tau^{I} H W_{\mu\nu}^{I} + h.c. 
\end{equation}
where, in the unitary gauge, the Higgs doublet $H$ reduces to its neutral component, which is $(v+h)/\sqrt{2}$ with $v=246$ GeV. This brings the sensitivity to the muon $g$-2 to NP scales of order $\Lambda \lesssim 100$ TeV (in the strongly-coupled regime). However, directly detecting new particles at such high scales is far beyond our current capabilities. Even if those particles were discovered, it would be nearly impossible to unambiguously link them to $\Delta a_\mu$ in a short time scale. 

The main idea is that a high-energy muon collider would offer a completely model-independent means to probe NP in the muon $g$-2. This is because the same dipole operator that generates the main contribution to $\Delta a_\mu$ unavoidably induces also a NP contribution to a muon scattering process, whose cross-section measurement will be equivalent to measuring $\Delta a_\mu$ within the effective approach.

In SMEFT, the relevant effective Lagrangian contributing to the muon $g$-2 up to one loop-order and after SSB, is 

\begin{equation} \label{eq:lag-smeft2}
    \mathcal{L}=\frac{C^\ell_{e\gamma}}{\Lambda^2} \frac{v}{\sqrt{2}}\bar{\ell}_L \sigma^{\mu\nu} e_R F_{\mu\nu} + \frac{C^\ell_{eZ}}{\Lambda^2} \frac{v}{\sqrt{2}} \bar{\ell}_L \sigma^{\mu\nu} e_R Z_{\mu\nu} + h.c. \,,
\end{equation}
where $C^\ell_{e\gamma}$ and $C^\ell_{e Z}$ are Wilson coefficients, and $\Lambda$ is the NP scale. 

Computing the Feynman diagrams that contribute to the $g$-2, one obtains 
\begin{equation}
    \Delta a_\ell \simeq \frac{4 m_\ell v}{\sqrt{2} e \Lambda^2} \left( C^\ell_{e\gamma} -\frac{3\alpha}{2\pi} \frac{{\rm cos}^2\theta_W-{\rm sin}^2\theta_W}{{\rm sin}\theta_W{\rm cos}\theta_W} \,C^\ell_{eZ} \,{\rm log}\frac{\Lambda}{m_Z} \right),
\end{equation}
where $\theta_W$ is the weak mixing angle.

\begin{figure}[h!]
     \centering
     \includegraphics[scale=.36]{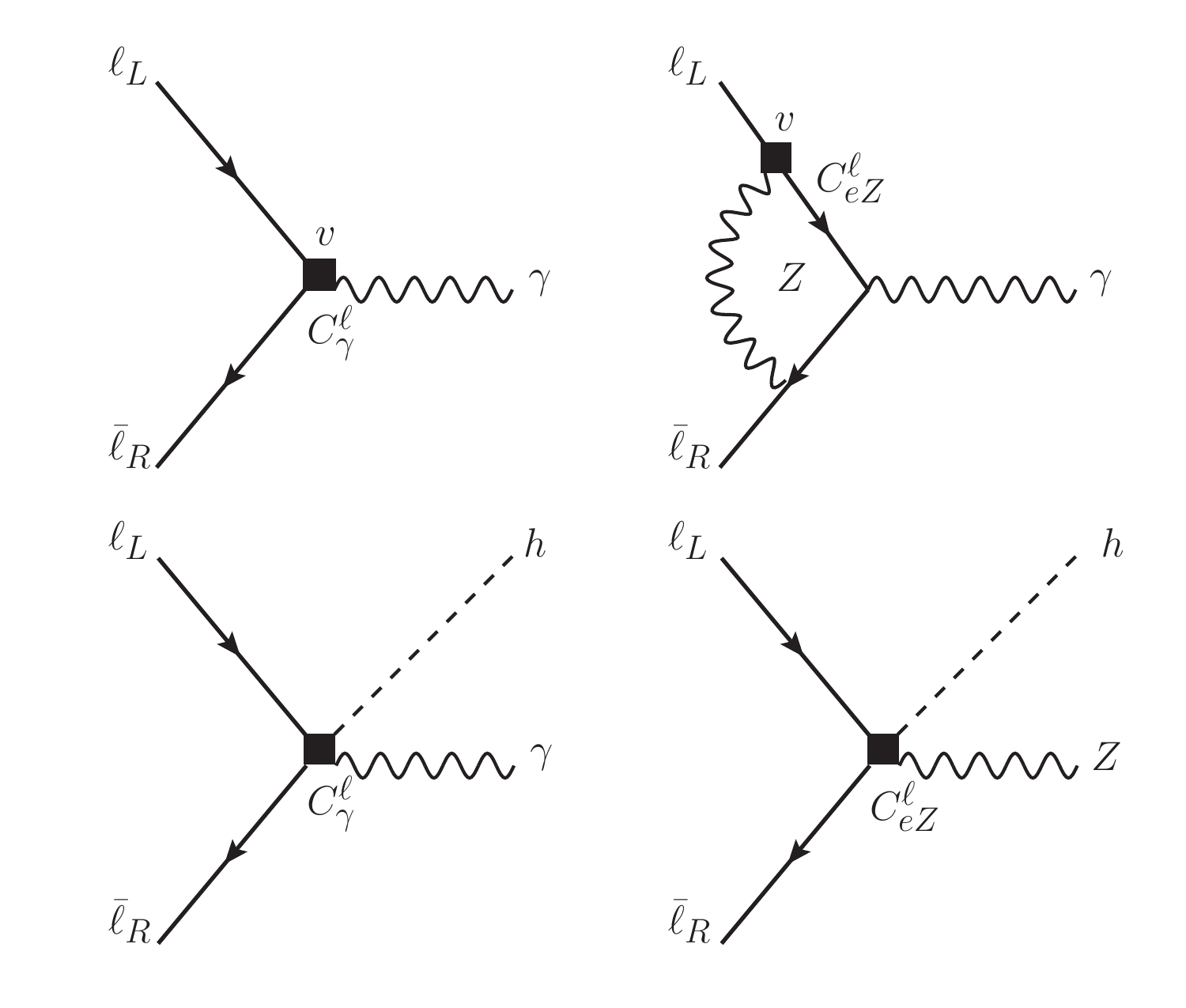}
     \caption{\textit{Upper row:} Feynman diagrams contributing to $(g\text{-2})_\ell$ in SMEFT, according to eq.~(\ref{eq:lag-smeft2}). \textit{Lower row:} Feynman diagrams of the corresponding $\ell^+\ell^-\to h+\gamma/Z$ scattering processes.}
     \label{fig:feynman}
 \end{figure}

But, as mentioned before, also high-energy scattering processes are generated, due to the Higgs field plus vacuum expectation value expansion after electroweak symmetry breaking, as can be seen explicitly in Fig.\ref{fig:feynman}. Then, from measuring the scattering events, we would be able to test the $g$-2 anomaly indirectly. 


 The key point is that, as both the $g$-2 contribution and the scattering process (either with the photon or with the $Z$ boson alone) are originated from the same effective operator, they share the same Wilson coefficient ---which is the only free parameter present in the analysis---. This direct correlation allows us to unambiguously link the two and extract low-energy information through high-energy measurements.

Finally, since the SMEFT Lagrangian, and thus the specific scattering processes, depend explicitly on the Higgs field, it is interesting to wonder if anything would be modified if the analysis is carried out using a HEFT approach. This HEFT effective analysis is more general (see e.g. refs.~\cite{Gomez-Ambrosio:2022qsi,Delgado:2023ynh} analyzing possible imprints of HEFT in multi-Higgs processes.). For instance, new dimension-6 contributions appear, that would be really suppressed in the SMEFT case (see ref. \cite{Gavela:2016bzc} for the power counting in HEFT) and the unambiguous relation between high and low energy processes could be lost ---due to the addition of unknown free parameters---. The explicit computation and main conclusions are derived in the following sections.

\section{HEFT Lagrangian}
\label{sec:3}
Following Ref.~\cite{Brivio_2016}, the HEFT operators with a single fermion current and up to two derivatives are contained in the following Lagrangian~\footnote{Higher-order HEFT Lagrangians are derived in Refs.~\cite{Sun:2022ssa,Sun:2022snw} and the EFT renormalization is studied in Refs.~\cite{Guo:2015isa,Buchalla:2017jlu,Alonso:2017tdy,Herrero:2021iqt}.}:
\begin{equation} \label{eq:heft-lag}
    \begin{split}
        \Delta\mathcal{L}_{2F}=&\sum_{j=1}^8 n_j^{\mathcal{Q}}\mathcal{N}_j^{\mathcal{Q}}+\sum_{j=9}^{28} \frac{1}{\Lambda}\left(n_j^{\mathcal{Q}}+i \tilde{n}_j^{\mathcal{Q}}\right)\mathcal{N}_j^{\mathcal{Q}} +\sum_{j=29}^{36} \frac{4\pi}{\Lambda}\left(n_j^{\mathcal{Q}}+i \tilde{n}_j^{\mathcal{Q}}\right)\mathcal{N}_j^{\mathcal{Q}}\\
        &+\sum_{j=1}^2 n_j^{\ell}\mathcal{N}_j^{\ell}+\sum_{j=3}^{11} \frac{1}{\Lambda}\left(n_j^{\ell}+i \tilde{n}_j^{\ell}\right)\mathcal{N}_j^{\ell} +\sum_{j=12}^{14} \frac{4\pi}{\Lambda}\left(n_j^{\ell}+i \tilde{n}_j^{\ell}\right)\mathcal{N}_j^{\ell}+h.c.,
    \end{split}
\end{equation}
where $n_j^{\mathcal{Q}}$, $n_j^{\ell}$, $\tilde{n}_j^{\mathcal{Q}}$ and $\tilde{n}_j^{\ell}$ are real coefficients smaller than unity~\footnote{We note the explicit $4\pi$ factors in the last of term of each line in eq.~(\ref{eq:heft-lag}), particularly in $\mathcal{N}_{12,13}^{\ell}$.} and the explicit form of all the operators can be found in Ref.~\cite{Brivio_2016}, along with enlightening explanations. In eq.~(\ref{eq:heft-lag}) the terms with two derivatives are suppressed by the NP scale
$\Lambda$, and the CP-even and -odd contributions have untilded and tilded coefficients, respectively.

Specifically, for the processes under study ($\mu^+ \mu^-\to h+\gamma/Z$ and $h\to\mu^+ \mu^- + \gamma/Z$), it is possible to prove that the only contributing operators are ($L_L$ is the SM lepton doublet and a right-handed neutrino does not appear in $L_R$
):
\begin{align}
    &\mathcal{N}_2^{\ell}(h)\equiv i\bar{L}_R \gamma_\mu U^{\dagger}\left\{V^\mu,T\right\}U L_R \mathcal{F},\nonumber
\\ &
    \mathcal{N}_4^{\ell}(h)\equiv \bar{L}_L \left\{V_\mu,T\right\}U L_R \partial^\mu \mathcal{F},\nonumber
\\ &
    \mathcal{N}_9^{\ell}(h)\equiv \bar{L}_L \sigma^{\mu\nu}\left\{V_\mu,T\right\}U L_R \partial_\nu \mathcal{F},\nonumber
\\ &
    \mathcal{N}_{12}^{\ell}(h)\equiv ig' \bar{L}_L \sigma^{\mu\nu} U L_R B_{\mu\nu} \mathcal{F},\nonumber
\\ &
    \mathcal{N}_{13}^{\ell}(h)\equiv ig \bar{L}_L \sigma^{\mu\nu} W_{\mu\nu}U L_R \mathcal{F},\label{eq:ops-heft}
\end{align}
with (below $L,R$ are $SU(2)_{L,R}$ transformations and $f_\pi$ the scale associated to the pseudo-Goldstone bosons, $V_\mu$ and $T$ transform in the adjoint of $SU(2)_L$ and the latter breaks custodial symmetry)
\begin{align}
   & U=\mathrm{exp}^{i\sigma_a\pi^a(x)/f_\pi},\;U(x) \to LU(x)R^\dagger,\nonumber\\
   & V_\mu\equiv(D_\mu U)U^{\dagger}, \quad T\equiv U\sigma_3 U^\dagger,\nonumber\\
   & D_\mu U(x)\equiv\partial_\mu U(x)+igW_\mu(x)U(x)-\frac{ig'}{2}B_\mu(x)U(x)\sigma_3 \nonumber\\
   & \mathcal{F}_i(h)=1+2a_i\frac{h}{v}+b_i\frac{h^2}{v^2}+...\label{eq:definitions}
\end{align}
%

The SM is reproduced if $a_0 = 1$, $b_0 = 1$ in the leading order term $(v^2/4) {\rm Tr}[(D_\mu U)^\dagger (D^\mu U)]$ $\mathcal{F}_0(h)$, the Higgs potential is $V(h)=v^4\left[\frac{m_h^2}{2v^2}\left(\frac{h}{v}\right)^3+\frac{m_h^2}{8v^2}\left(\frac{h}{v}\right)^4\right]$, and all other coefficients in operators with energy dimension larger than four vanish. A possible generalized function (with in principle unconstrained coefficients) in the Higgs potential does not play any role in this work.

The measured values of the $a_0$ and $b_0$ coefficients agree with SM predictions within uncertainties, but the current experimental accuracy of these measurements is only at the level of $\sim 10\%$. 
The unitarity condition for the HEFT amplitudes imposes a NP scale $\Lambda$ depending on the specific deviations from SM expectations. A significant departure implies a lower $\Lambda$, potentially around the TeV range. Conversely, for minor deviations, $\Lambda$ could be considerably higher, approaching arbitrarily large values in the limit where the SM is recovered. This relationship has been thoroughly explored from a geometric perspective, as elaborated in Ref.~\cite{Alonso:2015fsp}. For further details, we refer the reader to Ref.~\cite{Delgado:2014jda}, which discusses various processes and UV complete models, illustrating the operational mechanics of HEFT and its implications for determining the NP scale.

Certain processes, such as multi-boson radiation in the TeV range, could impact parameter sensitivity, particularly in scenarios where a low cutoff scale enhances these contributions. However, as noted in Ref.~\cite{Delgado:2014jda}, the effect could also lead to suppression relative to the naively expected \((E/4\pi v)^2\) factor, depending on the specific UV complete model. Given the large cutoff scale adopted in our analysis, this effect is not included in the main result of our calculation.

Then, if we use expressions (\ref{eq:definitions}) in (\ref{eq:ops-heft}), we obtain the following operators in the unitary gauge:
\begin{align} \label{eq:ops-Z}
    &\mathcal{N}_2^{\ell}(h)\equiv -g_z\bar{L}_R \gamma^\mu Z_\mu L_R \mathcal{F}_2,\nonumber
\\ &
    \mathcal{N}_4^{\ell}(h)\equiv ig_z\bar{L}_L Z_\mu L_R \partial^\mu \mathcal{F}_4,\nonumber
\\ &
    \mathcal{N}_9^{\ell}(h)\equiv ig_z\bar{L}_L \sigma^{\mu\nu}Z_\mu L_R \partial_\nu \mathcal{F}_9, 
\end{align}
with $g_z=e/(\sin{\theta_W}\cos{\theta_W})$ and where only the $Z_\mu$ part of the operator contributes \cite{Feruglio:1992wf}. The remaining $\mathcal{N}_{12}^{\ell}$ and $\mathcal{N}_{13}^{\ell}$ operators are analogous to the ones also appearing in the SMEFT dimension-six Lagrangian case, eq.~(\ref{eq:lag-smeft1}) (with a Higgs doublet instead of the $\mathcal{F}$ function, which can be different for every operator, as we emphasized with a subindex). Among these operators, only $\mathcal{N}_2^{\ell}$ is not suppressed by $1/\Lambda$ in eq.~(\ref{eq:heft-lag}), enhancing relatively its contributions at low energies.

Indeed, every HEFT operator can be generated within the SMEFT framework up to a certain dimension, considering the details discussed above. For completeness, Table \ref{tab:ops} presents the set of SMEFT operators that accounts for the additional HEFT terms.

\begin{table}[h]
\begin{center}
\renewcommand{\arraystretch}{1.4}
\begin{tabular}{cc}
\hline
\hline
HEFT & SMEFT (D=Dimension) \\
\hline
$\mathcal{N}_2^{\ell}$ & $\mathcal{Q}_{\varphi e}$ \cite{Grzadkowski:2010es} (D=6) \\
\hline
$\mathcal{N}_4^{\ell}$ & $Q^{(1),(2),(5)}_{leH^3D^2}$\cite{Murphy:2020rsh} (D=8)\\ 
\hline
$\mathcal{N}_9^{\ell}$ & $Q^{(3),(4),(6)}_{leH^3D^2}$\cite{Murphy:2020rsh} (D=8) \\
\hline
$\mathcal{N}_{12}^{\ell}$ & $\mathcal{Q}_{eW}$\cite{Grzadkowski:2010es}  (D=6)\\
\hline
$\mathcal{N}_{13}^{\ell}$ & $\mathcal{Q}_{eB}$\cite{Grzadkowski:2010es}  (D=6) \\
\hline
\hline
\end{tabular}
\caption{Correspondence between the HEFT operators analyzed in this work and the operators coming from SMEFT.} \label{tab:ops} 
\end{center}
\end{table}
It is also remarkable that none of these new HEFT operators ($ \mathcal{N}_2^{\ell}$, $\mathcal{N}_4^{\ell}$ and $ \mathcal{N}_9^{\ell}$) contribute to the anomalous magnetic moment of the muon directly, so only their contributions to the processes $\mu^+ \mu^-\to hZ$ and $h\to\mu^+ \mu^- Z$ will be taken into account.

Then, the corresponding HEFT Lagrangian for this work is given by
\begin{equation}
        \mathcal{L}= C_2^{\ell}\mathcal{N}_2^{\ell}+\frac{1}{\Lambda} C_4^{\ell}\mathcal{N}_4^{\ell}+\frac{1}{\Lambda} C_9^{\ell}\mathcal{N}_9^{\ell}+\frac{4\pi}{\Lambda}C_Z^{\ell}\mathcal{N}_Z^{\ell} +\frac{4\pi}{\Lambda}C_\gamma^\ell \mathcal{N}_\gamma^\ell,
\end{equation}
where the last two operators, $\mathcal{N}_Z^{\ell}$ and $\mathcal{N}_\gamma^{\ell}$ 
result from the linear combination of the $\mathcal{N}_{12}^{\ell}$ and $\mathcal{N}_{13}^{\ell}$ operators rendering gauge bosons in the mass eigenstate basis, and have the following form
%
\begin{align}
    \mathcal{N}_Z^\ell(h) &\equiv \bar{L}_L \sigma^{\mu\nu} L_R Z_{\mu\nu}\mathcal{F}_Z\,,\label{eq:Z-smeft}\\ 
    \mathcal{N}_\gamma^\ell(h) &\equiv \bar{L}_L \sigma^{\mu\nu} L_R F_{\mu\nu}\mathcal{F}_\gamma\,\label{eq:gamma-smeft}.
\end{align}

\section{HEFT $\mu^+\mu^-\to h\gamma$ cross section}
\label{sec:4}
In this section we compare the results obtained in the SMEFT and HEFT approaches, driven only by the $\mathcal{N}_\gamma^\ell$ operator in eq. (\ref{eq:gamma-smeft}), since all others contribute exclusively to the $Z$ scattering process, as we will derive in the following section. 

The differential and total cross section for the $\mu^+\mu^-\to h\gamma$ process, assuming that $\sqrt{s}\gg m_h$, where $\sqrt{s}$ is the collider center-of-mass energy, are given by ($\theta$ is the photon scattering angle)
\begin{align}
\frac{d\sigma^\text{\tiny{HEFT}}_{h\gamma}}{d\cos\theta}&=\frac{2\pi a_\gamma^2|C^\mu_\gamma|^2 s\ \text{sin}^2\theta}{ v^2\Lambda^2}\,,\nonumber\\
\sigma^\text{\tiny{HEFT}}_{h\gamma}&=\frac{8\pi a_\gamma^2|C^\mu_\gamma|^2 s}{3 v^2\Lambda^2}\,,
\end{align}
where one recovers the SMEFT expression by taking $a_\gamma=1$ and rescaling the HEFT Wilson coefficient:
\begin{equation}
    C^\mu_\gamma=\frac{v\, C_{e\gamma}^{\mu}}{8\pi\sqrt{2} \,\Lambda}\,,
\end{equation}
with $C_{e\gamma}^{\mu}$ the coefficient in the SMEFT Lagrangian in eq.~(\ref{eq:lag-smeft2}).

Note that the inclusion of $a_\gamma$ in the HEFT expression introduces an additional degree of freedom, which, without further constraints, breaks the direct and unambiguous link between low- and high-energy processes. In contrast, the SMEFT case, where $a_\gamma=1$ ---since the Higgs is a $SU(2)_L$ doublet---, allows for the direct relation previously discussed.

In order to compare the SMEFT and HEFT approaches, we follow the method used in Ref.~\cite{Buttazzo:2020ibd}, where the authors considered a cut-and-count experiment in the $b\bar{b}$ final state ($h\to b\bar{b}$), which has the highest signal yield (with branching ratios $\mathcal{B}(h\to b\bar{b})=0.58$ and $\mathcal{B}(Z\to b\bar{b}=0.15)$). The significance of the signal is defined as $N_S/\sqrt{N_S+N_B}$, with $N_S(N_B)$ the number of signal (background) events. Thus, requiring at least one jet to be tagged as a $b$, and assuming a plausible b-tagging efficiency, $\epsilon_b=80\%$, one gets the $95\%$ C.L. reach on the muon anomalous magnetic moment $\Delta a_\mu$, as a function of the collider center-of-mass energy, $\sqrt{s}$, from the process $\mu^+\mu^-\to h\gamma$. See Ref.~\cite{Buttazzo:2020ibd} for all the details in the SMEFT derivation of this result.

We stress here that the SM irreducible $\mu^+\mu^-\to h\gamma$ background can be neglected for $\sqrt{s}>1$ TeV, being the $Z\gamma$ events the main source of contamination, where the $Z$ boson is incorrectly reconstructed as a Higgs, as commented in Ref.~\cite{Buttazzo:2020ibd}. For this specific channel we take the central region cut $|\cos{\theta}|\leq 0.6$, where the significance of the signal is maximized, leading to the following signal and background total decay rates at $\sqrt{s}=30$ TeV:
\begin{equation}
  \sigma_{h\gamma}^{cut}\approx 0.40\,\text{ab} \left(\frac{\Delta a_\mu}{3\times10^{-9}} \right)^2 a_{\gamma}^2,\quad\quad  \sigma_{Z\gamma}^{cut}\approx 80.5\,\text{ab}, 
\end{equation}
where ab denotes attobarns.

After the Wilson coefficient rescaling, the only difference between both approaches is due to the ``$a_\gamma$'' parameter that appears in the HEFT $\mathcal{F}$-function expansion. Following the discussion and limits presented in Ref.\cite{Gomez-Ambrosio:2022qsi}, in the following examples we analyze cases where $\frac{1}{5}< a_i\leq5$, as a first reasonable approach.

\begin{figure}[h!]
     \centering
     \includegraphics[scale=.75]{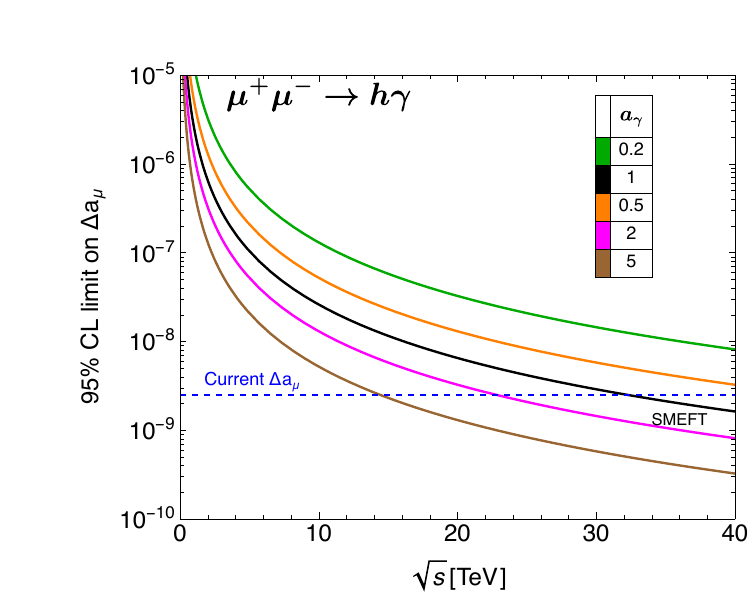}
     \caption{95$\%$ C.L. reach on the muon anomalous magnetic moment $\Delta a_\mu$ for operator $\mathcal{N}_\gamma^{\ell}$ contribution with different $a_\gamma$ values. In this and later figures, the black curve shows the SMEFT result, corresponding to $a_\gamma=1$. The dashed line shown as `Current $\Delta a_\mu$' displays the 2020 White Paper prediction \cite{Aoyama:2020ynm}, to be updated soon.}
     \label{fig:CgammaHEFT}
 \end{figure}

In fig. \ref{fig:CgammaHEFT}, we show the $95\%$ C.L. reach from $\mu^+\mu^-\to h\gamma$ on $\Delta a_\mu$ as a function of the collider energy, assuming that only $\mathcal{N}_\gamma^{\ell}$ contributes to $\Delta a_\mu$. The point where the curves and the dashed line intersect shows the specific energy needed to test the presumed NP under the assumption that it generates a contribution of $2.49\times 10^{-9}$ to the $\Delta a_\mu$. In the same way, the region below the dashed line is interesting under the assumption that this NP generates only a partial contribution to the current $\Delta a_\mu$, which would be equivalent to lowering the dashed line (in case the revised $a_\mu$ turns out to be smaller than in ref.~\cite{Aoyama:2020ynm}). The black line reproduces the SMEFT result, corresponding to $a_\gamma=1$, and we also show the curves generated when $a_\gamma=0.2, 0.5, 2$ and $5$.


  Note that for values of $a_\gamma>1$, the sensitivity to test the $g$-2 anomaly improves with respect to that in SMEFT. For instance, if we take $a_\gamma=5$, the $95\%$ C.L. on $\Delta a_\mu$ is reached at a center-of-mass energy of $\sim 14$ TeV. In SMEFT, a $\sqrt{s}\sim 30$ TeV is needed to test the $g$-2 anomaly with the desired accuracy. This behavior was already expected, since increasing $a_\gamma$ means a higher number of signal events. 

We want to emphasize that for the SMEFT case, under the assumption that the NP generates a contribution of $2.49\times 10^{-9}$ to the $\Delta a_\mu$, the required muon collider energy is completely fixed in order to generate the desired signal events expected from the unambiguous relation between the low- and high-energy processes. That is why only one SMEFT curve is present in Fig. \ref{fig:CgammaHEFT}. In other words, if the NP contribution to $\Delta a_\mu$ is assumed to really exist, a fixed number of scattering events, generated by the same NP, can be tested at a muon collider with a specific center-of-mass energy, without requiring additional assumptions.

In the HEFT analysis, the above discussion is no longer true, since having an extra degree of freedom $a_\gamma$ modifies the relation between the different processes. Then, it could happen that even if the same NP contribution to $\Delta a_\mu$ is considered, we may not see enough high-energy scattering events ($|a_\gamma|<1$) or we could get a higher event rate at smaller collider energies ($|a_\gamma|>1$), as shown in Fig. \ref{fig:CgammaHEFT}. Consequently, the HEFT conclusion will depend on the specific $a_\gamma$ value.

On the other hand, from a phenomenological perspective, $a_\gamma$ is expected to be close to the SM prediction, $a_\gamma \approx 1$. Regardless, this analysis highlights the need for independent measurements to constrain the value of $a_\gamma$. This would ensure an unambiguous relationship between low- and high-energy processes, even within the HEFT framework, allowing for a definitive conclusion.

\section{HEFT $\mu^+ \mu^-\to hZ$ cross section}
\label{sec:5}
Using the HEFT Lagrangian, the differential cross sections for the process $\mu^+ \mu^-\to hZ$ at $\sqrt{s}\gg m_h$ are (we first give the contributions from squaring the amplitude for each operator and then the only relevant interference):
    \begin{equation}
    \begin{split}
    & \bullet\ \text{Operator}\ \mathcal{N}_2^{\mu}:\ C^2_{2}  \frac{s\,{\rm sin}^2\theta}{512\pi m_Z^2}\,. \quad\quad \bullet\ \text{Operator}\ \mathcal{N}_4^{\mu}:\   \frac{|C_{4}|^2}{\Lambda^2}  \frac{s^2}{256\pi m_Z^2}\,. \\
    & \bullet\ \text{Operator}\ \mathcal{N}_9^{\mu}:\  \frac{|C_{9}|^2}{\Lambda^2}  \frac{s^2\,{\rm cos}^2\theta}{256\pi m_Z^2}\,. \quad\quad \bullet\ \text{Operator}\ \mathcal{N}_Z^{\mu}:\ \frac{|C_{Z}|^2}{\Lambda^2}  \frac{s\,{\rm sin}^2\theta}{32\pi}\,. \\
    &\bullet\ \text{Interference}\ \mathcal{N}_9^{\mu}-\mathcal{N}_Z^{\mu}:\ -\frac{\operatorname{Re}\left(C_9^* C_{Z}\right)}{\Lambda^2}  \frac{s\,{\rm cos}^2\theta}{32\pi}\,.
      \end{split}
    \end{equation}
This leads to
\begin{equation} \label{eq:cross-section}
    \begin{split}
        &\frac{d\sigma_{\text{\tiny{HEFT}}}}{d\cos\theta}=\frac{C^2_{2} s\,{\rm sin}^2\theta}{512\pi m_Z^2}+\frac{s}{\Lambda^2}\left[ \frac{|C_{Z}|^2 \,{\rm sin}^2\theta}{32\pi} + \frac{|C_{4}|^2 s}{256\pi m_Z^2}+ \frac{|C_{9}|^2 s\,{\rm cos}^2\theta}{256\pi m_Z^2} -\frac{\operatorname{Re}\left(C_9^* C_{Z}\right)  \cos^2\theta}{32\pi}\right],\\
        &\sigma_{\text{\tiny{HEFT}}}=\frac{C^2_{2} s}{384\pi m_Z^2}+\frac{s}{24\pi \Lambda^2}\left[ |C_{Z}|^2 + \frac{3|C_{4}|^2 s}{16 m_Z^2}+ \frac{|C_{9}|^2 s}{16 m_Z^2}-\frac{\operatorname{Re}\left(C_9^* C_{Z}\right)}{2}\right],
    \end{split}
\end{equation}
where, for simplicity, we have redefined the coefficients as follows 
\begin{equation} \label{eq:coefficients}
    C_{Z}\to \frac{2(4\pi)a_{Z}}{v} C^\mu_{Z}, ~C_9\to\frac{2ia_9}{v} C^\mu_9 g_z, ~C_2\to\frac{-2a_2}{v} C^\mu_2 g_z,  ~C_4\to\frac{2ia_4}{v} C^\mu_4 g_z.
\end{equation}

Note that the cross section in eq. (\ref{eq:cross-section}) has the expected energy dimensions after the substitution of the coefficients in eq. (\ref{eq:coefficients}). Furthermore, as shown explicitly in eq.~(\ref{eq:cross-section}) for the new HEFT operators, due to the specific energy dependence (where $m_h\ll\sqrt{s}\ll\Lambda$) and its numerical coefficient, the $\mathcal{N}_2$ operator will have the largest contribution, followed by $\mathcal{N}_4$ and $\mathcal{N}_9$ respectively, as will be discussed in detail later.

The interference between operators $\mathcal{N}_2^{\ell}$-$\mathcal{N}_Z^{\ell}$, $\mathcal{N}_2^{\ell}$-$\mathcal{N}_9^{\ell}$ and $\mathcal{N}_2^{\ell}$-$\mathcal{N}_4^{\ell}$ are all proportional to the lepton mass and are therefore negligible in the energy range of interest ($\sqrt{s}\gg m_h$), while the interferences between operators $\mathcal{N}_4^{\ell}$-$\mathcal{N}_Z^{\ell}$ and $\mathcal{N}_4^{\ell}$-$\mathcal{N}_9^{\ell}$ are identically zero. Also, the result obtained from the operator $\mathcal{N}_Z^{\ell}$ alone has the same functional form as the one obtained in the SMEFT case \cite{Buttazzo:2020ibd}, both related by a rescaling of the Wilson coefficient, $C_Z=\frac{ C^{\mu}_{eZ}}{ \sqrt{2} \Lambda}$ (where $C^{\mu}_{eZ}$ is the coefficient in the SMEFT Lagrangian in eq.~(\ref{eq:lag-smeft2})).

We would like to stress that the calculation for the cross sections presented in this paper includes diagrams at tree level only, i.e., at leading-order (LO). For a more precise estimation of our results, one should include next-to-leading order (NLO) corrections. To address this issue, we refer to Ref. \cite{Bredt:2022dmm}, which provides a complete analysis of NLO electroweak (EW) corrections and initial-state radiation (ISR) effects in multiple massive boson production processes at a future muon collider. Specifically, for the process $\mu^+\mu^-\to H Z$, they concluded that the overall effect is a decrease of the cross section of $\sim 20\%$ (with respect to the LO result) at $\sqrt{s}\sim 16$ TeV, and an even larger decrease at higher energies. It is also important to note that these results were obtained in a fully inclusive analysis. Therefore, although the general effect should apply approximately to our exclusive calculation, $\mu^+\mu^-\to HZ\to (\bar{b}b)(\bar{b}b)$, a difference in the magnitude of such effect could be expected. The previous discussion is meant to stand out the importance of the NLO corrections for the precise computation of a cross section at high energies.

However, such a detailed estimation lies beyond the scope of the current study, and we would like to emphasize that our goal of showing that the HEFT framework is more appropriate than SMEFT to address the search of new physics related to the $g$-2 in a muon collider \cite{Buttazzo:2020ibd}, still stands. As mentioned before, the current experimental accuracy that would confirm the way in which the three Goldstone bosons and the massive scalar Higgs are embedded in the SM, is only at the level of $\sim 10\%$. On the other hand, from the results obtained in this analysis, even at LO, we may already highlight the possible difference in the $95\%$ C.L. reach on the muon anomalous magnetic moment $\Delta a_\mu$, depending on the effective approach.

Now we can give a rough estimation of the new HEFT contributions compared with the SMEFT one. Following the same cut-and-count experiment method outlined previously, together with hadronic decays of the $Z$ ($\mathcal{B}(Z\to {\rm had})=0.699$), one gets the 95$\%$ C.L. reach on the muon anomalous magnetic moment $\Delta a_\mu$, as a function of the collider center-of-mass energy $\sqrt{s}$, from the process $\mu^+ \mu^-\to hZ$.

Since our main goal is to study the impact of the new HEFT contributions to the possible reach on the muon anomalous magnetic moment $\Delta a_\mu$, we have to be careful when defining the signal and background events. As done in Ref.~\cite{Buttazzo:2020ibd}, the significance of the signal is defined as $N_S/\sqrt{N_T}$, where $N_T$ is the total number of events generated including the possible background $N_T=N_S+N_B$. In this case, the signal events will be due to all the operators that generate $\mu^+ \mu^-\to hZ$ but also contribute to $\Delta a_\mu$, where we can directly relate the Wilson coefficient entering the scattering process with the $\Delta a_\mu$ value. All other contributions, including the SM and new HEFT operators, that generate $\mu^+ \mu^-\to hZ$ but do not contribute directly to $\Delta a_\mu$, will be considered as background for the sought signal. 

Then, as already discussed, the $\mathcal{N}_2^{\ell}$, $\mathcal{N}_4^{\ell}$ and $\mathcal{N}_9^{\ell}$ operators, which do not contribute to $\Delta a_\mu$, must be considered as part of the ``background'' events in order to study the reach from $\mu^+\mu^-\to hZ$ on $\Delta a_\mu$ as a function of the collider energy at a $95\%$ C.L. 
This is a new challenge compared to the $\mu^+\mu^-\to h\gamma$ case, as HEFT background contributions depend on unknown $a_i$ parameters and Wilson coefficients and therefore the subtraction of these backgrounds would require knowledge of such parameters.

Finally, we assumed that only $\mathcal{N}_Z^{\ell}$ contributes to $\Delta a_\mu$ \footnote{This might correspond to an unnatural scenario but is nevertheless meaningful, as explained in Ref.~\cite{Buttazzo:2020ibd}.} and considered the SM irreducible background to be:
\begin{equation}
    \sigma_{Zh}^{SM}\approx 122\,\text{ab}\left(\frac{10\, \text{TeV}}{\sqrt{s}} \right)^2.
\end{equation}
The bound on $\Delta a_\mu$ (at a $95\%$ C.L. reach), that could be extracted from the high-energy measurement, in terms of all the HEFT Wilson coefficients, after fixing $m_Z$, $v$, $g_Z$ and $\Lambda\approx 100$ TeV, is given by:
\begin{align}
    \Delta a_\mu\approx& \frac{10^{-8}}{a_Z} \Bigg\{ 16 a_9 {\rm Im}(C_9^\mu)+3.4 \bigg[ \frac{1.1}{s^2}+21 a_9^2({\rm Im}(C_9^\mu))^2+ \frac{12}{s} \Big(\frac{3.4}{s^2} + 3.7\times 10^5 a_2^2 |C_2^\mu|^2\nonumber\\ & +s \big\{1.1\times 10^2 a_4^2 |C_4^\mu|^2 + 37 a_9^2 |C_9^\mu|^2\big\} \Big)^{1/2} \bigg]^{1/2} \Bigg\},
    \label{eqn:amu}
\end{align}
where the energy units of each term can be traced following the $s$ (TeV$^2$) powers.

From a quick inspection of eq.~(\ref{eqn:amu}) we can advance some properties of the new HEFT terms. As an example, the numerical coefficient indicates that the $\mathcal{N}_2$ operator could generate the largest contribution in the energy range of interest, as anticipated from eq. (\ref{eq:cross-section}). Also, the explicit global dependence on $a_Z$ suggests that the sensitivity to a small $\Delta a_\mu$ will be achieved more quickly at a specific $\sqrt{s}$ for a larger value of $a_Z$, as already expected. All these properties will be discussed next. 

\begin{figure}[h!]
     \centering
     \includegraphics[scale=.75]{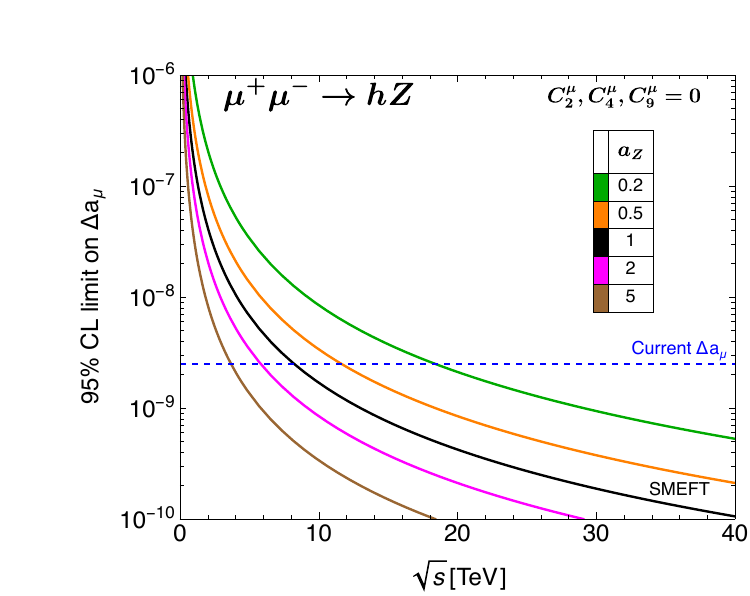}
     \caption{95$\%$ C.L. reach on the muon anomalous magnetic moment $\Delta a_\mu$ for operator $\mathcal{N}_Z^{\ell}$ contribution with different $a_Z$ values. The black curve shows the SMEFT result, corresponding to $a_Z=1$.}
     \label{CZHEFT}
 \end{figure}

This can be done systematically considering the new HEFT contributions one-by-one. As a first result, we work only with the $\mathcal{N}_Z^{\ell}$ operator, which is the same appearing in the SMEFT case. This scenario is analogous to the comparison between SMEFT and HEFT approaches using the $\mathcal{N}^\ell_\gamma$ operator, where after the Wilson coefficient rescaling ($C_Z=\frac{ C^{\mu}_{\text{eZ}}}{ \sqrt{2} \Lambda }$), the only difference between both frameworks is now due to the ``$a_Z$'' parameter that appears in the HEFT $\mathcal{F}$-function expansion. To study its possible effects, we plot the 95$\%$ C.L. reach on $\Delta a_\mu$ in Fig.~\ref{CZHEFT}, considering different ``$a_Z$'' values, being $a_Z=1$ identical to the SMEFT case, shown as the black curve. 
 
 As we can see, in an optimistic scenario, we find that a value of $\Delta a_\mu =2.49\times 10^{-9}$ can be tested at $95\%$ C.L. at a 4-7 TeV collider due to the presence of the extra ``$a_Z$'' parameter in the HEFT case, instead of the 10 TeV required for the SMEFT analysis. However, as discussed in the last section, this cannot be used as a model-independent probe of the muon $g$-2 in this context without additional information about $a_Z$.

 Now we can study the effects of the new $\mathcal{N}_2^{\mu}$, $\mathcal{N}_4^{\mu}$ and $\mathcal{N}_9^{\mu}$ operators one by one, together with the $\mathcal{N}_Z^{\mu}$ contribution, in order to compare it with the SMEFT case. For this purpose we focus on a new single HEFT contribution at a time, setting the other Wilson coefficients to zero. After that, we plot the 95$\%$ C.L. reach on $\Delta a_\mu$ considering reasonable values of the corresponding ``$a_i$'' parameter and $C_i^\mu$ Wilson coefficient.
 
 Motivated by the Wilson coefficient rescaling ($C^\mu_Z=\frac{v\ C^\mu_{eZ}}{2(4\pi)\sqrt{2}\Lambda}$), an $\mathcal{O}(1)$ SMEFT Wilson coefficient will imply an $\mathcal{O}(10^{-4}-10^{-5})$ HEFT coefficient \footnote{This agrees with the current $C^\mu_{2}$ upper-limit~\cite{Eboli:2021unw}.} and it would be naturally expected that the other HEFT couplings will be even more suppressed since their counterparts in SMEFT appear as higher dimension operators. Then, even if it is true that the HEFT coefficients are not restricted to have these values, for a first and reasonable approach we shall study the HEFT operators effects considering their Wilson coefficients varying in the range $10^{-5}<C^\mu_{i}<10^{-7}$. This assumption, again, is necessary due to the additional degrees of freedom, and precise information about $C^\mu_{i}$ would be crucial for a definite conclusion.

 In all the following examples we try to show some limiting cases, for illustrative purposes, where in the left figure we fix the Wilson coefficient and plot the effects of the $a_i$ and $a_Z$ parameters considering their minimum and maximum values. In the right figure the procedure is essentially analogous, fixing $a_i=1$ in order to see the interplay between the Wilson coefficient and $a_Z$ for their extreme values. In this way, almost all possible cases will be contained in the intermediate region delimited by the outer curves of each graph.

\begin{figure}[h!]
\begin{subfigure}[]{0.5\textwidth}
   \includegraphics[width=1\linewidth]{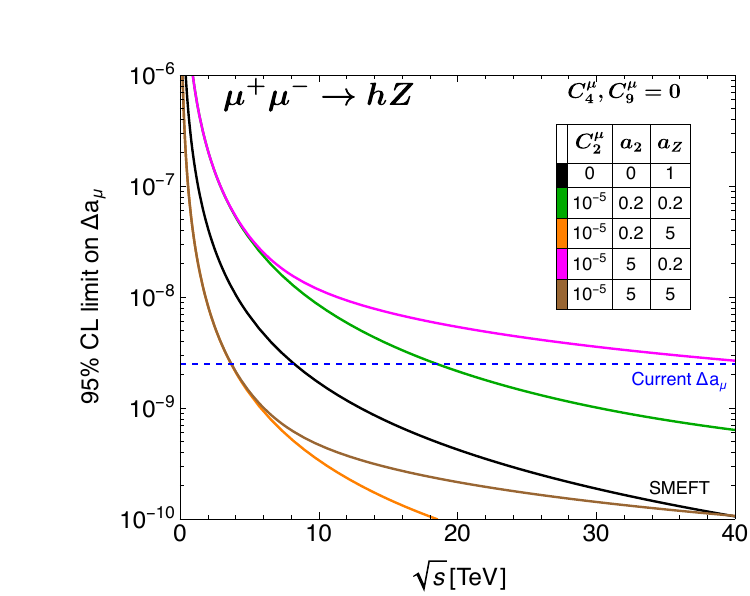}
   \caption{\footnotesize{$a_2$ effect for a fixed $C_2^\mu$ Wilson coefficient.}}
\end{subfigure}
\begin{subfigure}[]{0.5\textwidth}
   \includegraphics[width=1\linewidth]{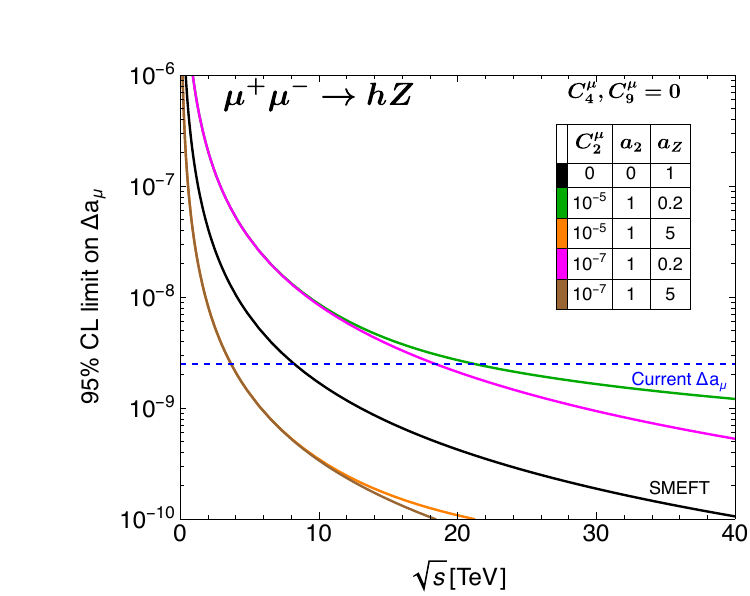}
   \caption{\footnotesize{$C_2^\mu$ Wilson coefficient effect for a fixed $a_2$ value.}}
\end{subfigure} 
\caption{95$\%$ C.L. reach on the muon anomalous magnetic moment $\Delta a_\mu$ with the $\mathcal{N}_2^{\ell}$ operator contribution. The black curve shows the SMEFT result.}
 \label{fig:C2HEFT}
 \end{figure}

For the $\mathcal{N}_2^{\mu}$ operator we get Fig.~\ref{fig:C2HEFT}, being the black curve the SMEFT case.

 Strictly speaking, one could consider operator $\mathcal{N}_2^\mu$ as part of the SMEFT approach, up to dimension 6 (as shown in table \ref{tab:ops}). However, incorporating its contribution into the black line in Fig. \ref{fig:C2HEFT} ---for all reasonable values of the Wilson coefficient $C_2^\mu$---, does not alter the intersection point between the SMEFT curve (black curve) and the current $\Delta a_\mu$ value (dashed blue line). Therefore, for clarity and simplicity, only a single SMEFT curve (that does not include $\mathcal{N}_2^\mu$) is presented in this case without affecting the overall conclusions.

 We find that even though the $\mathcal{N}_2^{\mu}$ operator would distort the SMEFT case in almost all scenarios, only a couple of them would be useful to test $\Delta a_\mu =2.49\times 10^{-9}$ at a smaller collider energy than that required within the SMEFT approach. Specifically, for a maximum value of $a_Z$, the collider energy needed could be about 4 TeV, approximately. We also see that smaller $C_2^\mu$ and $a_2$ values could help to increase the $\Delta a_\mu$ sensitivity for a given $a_Z>1$ value. 

 This result can be easily understood, since a larger $a_Z$ value together with smaller $C_2^\mu$ and $a_2$ contributions will generate more signal events while decreasing the background ones, respectively, achieving the desired sensitivity at a lower energy scale. If $a_Z<1$, the signal events are reduced, and if $C_2^\mu$ or $a_2$ are large, the background increases; in both cases losing sensitivity to $\Delta a_\mu$, as shown in the plots. Again, the previous discussion holds only if all new HEFT degrees of freedom are well-known, allowing for an unambiguous characterization of HEFT background and signal events.

\begin{figure}[h!]
\begin{subfigure}[]{0.5\textwidth}
   \includegraphics[width=1\linewidth]{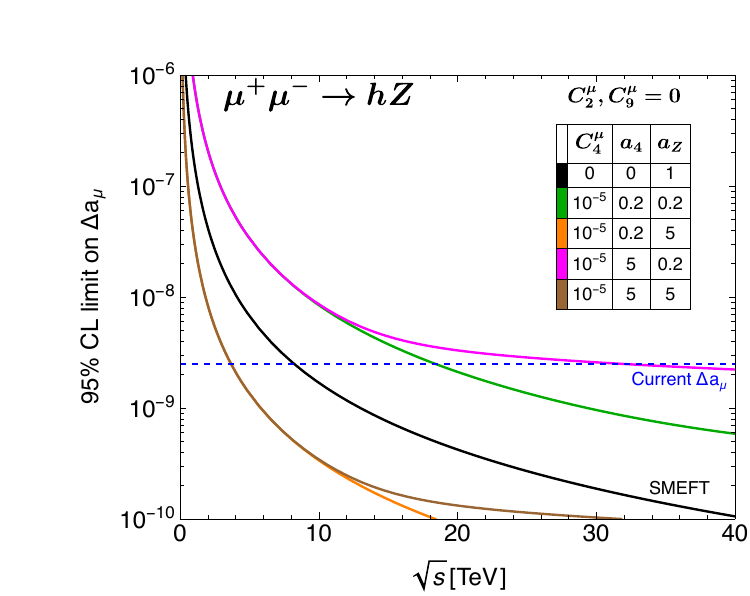}
   \caption{\footnotesize{$a_4$ effect for a fixed $C_4^\mu$ Wilson coefficient.}}
\end{subfigure}
\begin{subfigure}[]{0.5\textwidth}
   \includegraphics[width=1\linewidth]{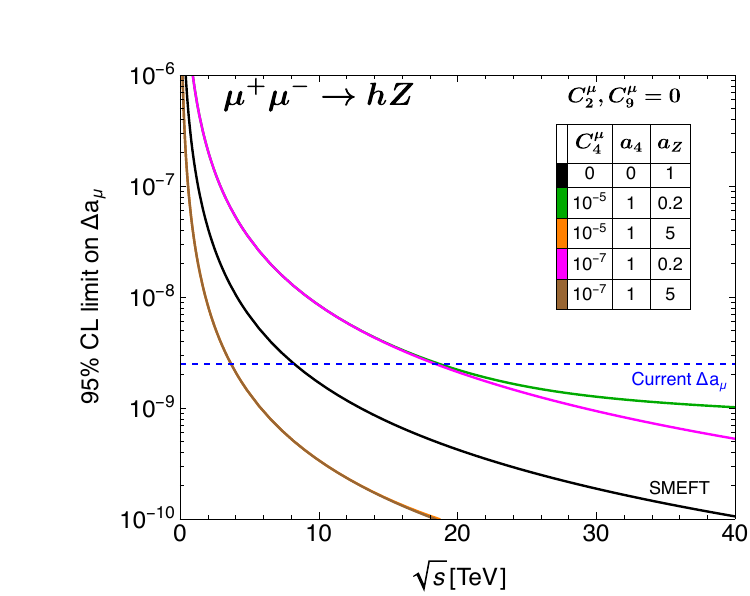}
   \caption{\footnotesize{$C_4^\mu$ Wilson coefficient effect for a fixed $a_4$ value. }}
\end{subfigure}
 \caption{95$\%$ C.L. reach on the muon anomalous magnetic moment $\Delta a_\mu$ for the operator $\mathcal{N}_4^{\ell}$ contribution. The black curve shows the SMEFT result.}
 \label{fig:C4HEFT}
 \end{figure}

\begin{figure}[h!]
\begin{subfigure}[]{0.5\textwidth}
   \includegraphics[width=1\linewidth]{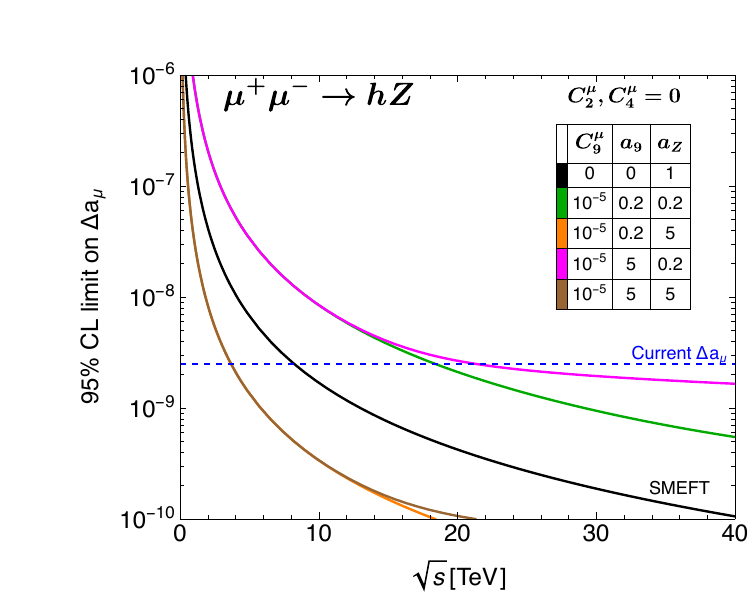}
   \caption{\footnotesize{$a_9$ effect for a fixed $C_9^\mu$ Wilson coefficient.}}
\end{subfigure}
\begin{subfigure}[]{0.5\textwidth}
   \includegraphics[width=1\linewidth]{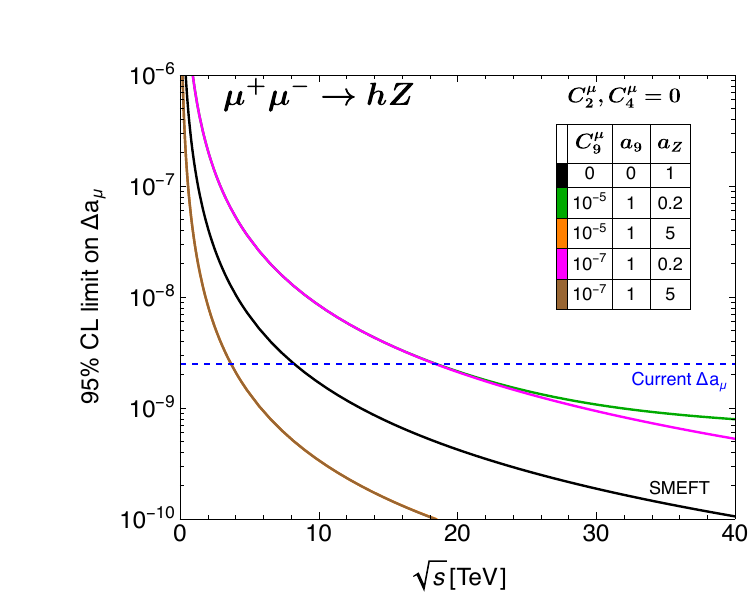}
   \caption{\footnotesize{$C_9^\mu$ Wilson coefficient effect for a fixed $a_9$ value.}}
\end{subfigure}
 \caption{95$\%$ C.L. reach on the muon anomalous magnetic moment $\Delta a_\mu$ for operator $\mathcal{N}_9^{\ell}$ contribution. The black curve shows the SMEFT result.}
 \label{fig:C9HEFT}
 \end{figure}

We keep going with the next HEFT operator, $\mathcal{N}_4^{\mu}$, whose contribution is shown in Fig.~\ref{fig:C4HEFT}, again for both cases already explained.
  The results are basically the same, as we can see, the scenarios with larger $a_Z$ values could explore $\Delta a_\mu=2.49\times 10^{-9}$ at a 4 TeV collider energy, while other combinations with $a_Z<1$ and large ($C_4^\mu$, $a_4$) values would lose sensitivity to $\Delta a_\mu$ at reduced energies. It is also interesting that there are small differences between the $\mathcal{N}_2^{\ell}$ and $\mathcal{N}_4^{\ell}$ contribution, as can be seen directly from the plots, but they do not interfere with the conclusions of this work. Finally, the same analysis for the last $\mathcal{N}_9^{\mu}$ operator contribution is displayed in Fig.~\ref{fig:C9HEFT}, leading to identical conclusions as above.

\begin{figure}[h!]
\begin{subfigure}[]{0.5\textwidth}
   \includegraphics[width=1\linewidth]{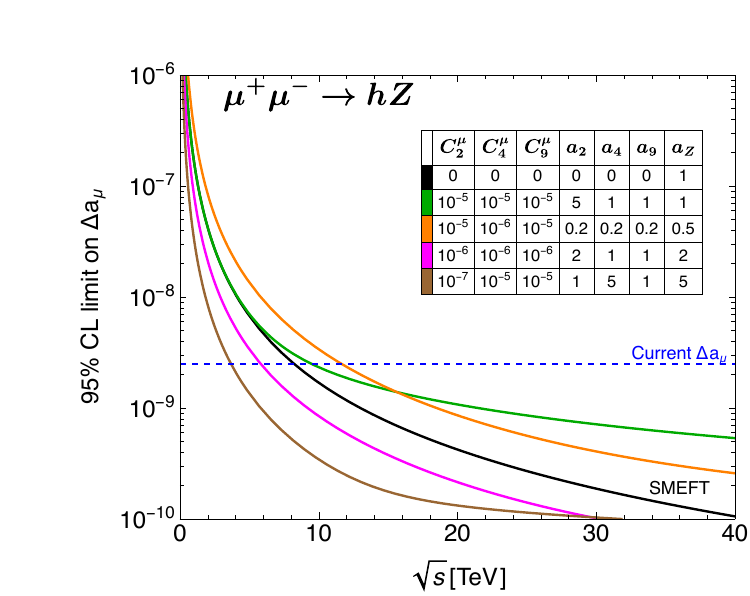}
\end{subfigure}
\begin{subfigure}[]{0.5\textwidth}
   \includegraphics[width=1\linewidth]{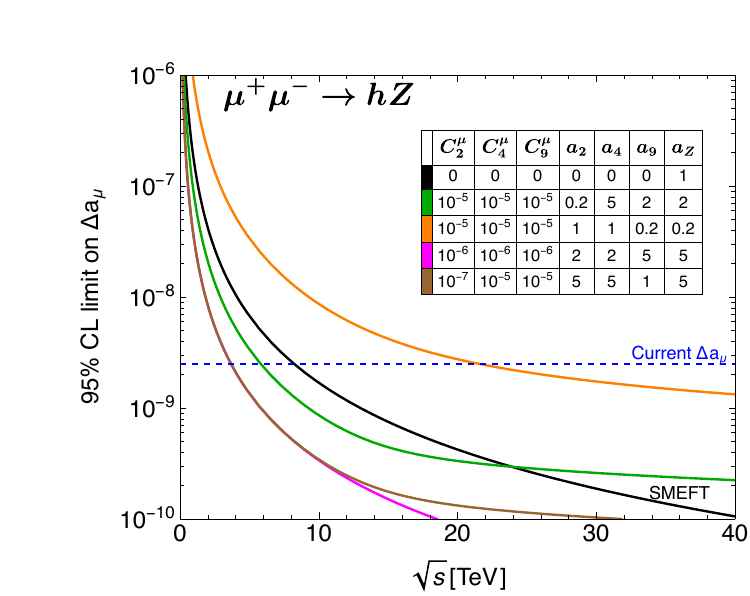}
\end{subfigure}
 \caption{95$\%$ C.L. reach on the muon anomalous magnetic moment $\Delta a_\mu$ for all dimension-6 HEFT operators contributions for different sets of $C_i^\mu$ Wilson coefficients and $a_i$ values.}
 \label{CZTHEFT}
 \end{figure}

For a last discussion we plot a set of different general cases, considering many $C_i^\mu$ and $a_i$ values in order to analyze the modification due to the entire HEFT contribution, shown in Fig.~\ref{CZTHEFT}.

 Here it is interesting to note the different behaviors that the HEFT case could induce depending on all the coupling values. Some of them would require a higher center of mass energy to test the $g$-2 anomaly at $95\%$ C.L. However, there are possible scenarios where the collider energy required is smaller, as shown in Fig.~\ref{CZTHEFT}, that could test the muon anomalous magnetic moment at $\sim 4-7$ TeV collider energy.

Another interesting remark is that, in some cases, one specific configuration of Wilson coefficients and $a_i$ values could lead to a higher sensitivity to $\Delta a_\mu$ in some energy region but could be worse in other energy range. This is explicitly shown in Fig.~\ref{CZTHEFT}, by the green and orange curves on the left-hand side or by the green and black ones on the right-hand side. All this shows the different possible behaviors in the general case, agreeing with all the conclusions previously given. 

We stress that the results shown in Fig.~\ref{CZTHEFT} have reasonable values for the $C_i^\mu$ Wilson coefficient and $a_i$ parameters. However, experimental data allowing us to constrain them are required in order to develop a more precise analysis. Specifically for the case of $hZ$ production, where several new operators can contribute within the HEFT approach, without affecting the $g$-2.

\section{Conclusions}
The discrepancy between the theoretical and experimental value of the muon $g$-2 has been tantalising for several years now, as one of the most promising hints for new physics.
Following Ref. \cite{Buttazzo:2020ibd} --which focused on the SMEFT analysis-- in this work we explore, within the HEFT formalism, the possibility to test this discrepancy by measuring the $\mu^+\mu^-\to h+\gamma/Z$ cross sections in a muon collider. 

We have computed the $\mu^+\mu^-\to h+\gamma/Z$ cross sections  and the $h\to\mu^+\mu^-+\gamma/Z$ rare decays (appendix \ref{App:HDecay}), using the dimension six effective operators within the HEFT approach. Some of these also generate NP contributions to the lepton dipole moments, therefore the NP that contributes to the muon $g$-2 could be probed indirectly. In this way we could test, in a model-independent way, a precise low-energy prediction of the SM using the tools of the high-energy physics frontier, as long as we had precise knowledge of the HEFT parameters.  

The HEFT approach provides a more general framework than SMEFT to search for NP in the Higgs sector, and could lead to distinct phenomenological scenarios. For instance, the aforementioned deviation between the theoretical and experimental values for the muon $g$-2, that in principle could be tested in $\mu^+\mu^-\to h+\gamma/Z$ within SMEFT in an unambiguous way, cannot be used as a model-independent probe of the muon $g$-2 in the HEFT context (due to the new degrees of freedom).

This is due to the singlet nature of the Higgs in the HEFT, where all interactions involving the physical Higgs boson have independent couplings $a_i$ that cannot be directly related to quantities without the Higgs. Therefore, without precise information about the new degrees of freedom, observing a deviation at high energies would not necessarily confirm a NP contribution to $\Delta a_\mu$, and the absence of a deviation would not definitively rule it out.

All of these discussions are specifically important for the $\mu^+\mu^-\to hZ$ process, where the NP HEFT backgrounds contributions could not be subtracted without further knowledge of the unknown parameters. Therefore, complementary measurements are crucial to determine the values of the relevant parameters. This motivates further theoretical and experimental analysis.

Finally, a more realistic analysis would require a more sophisticated inclusive signal and background definition, which serves as motivation for future work, including a dedicated discussion on the LO and NLO corrections, especially in light of future data from a prospective muon-collider.

\section*{Acknowledgements}
It is our pleasure to thank the very useful advices and observations by Juanjo Sanz Cillero and José Wudka on the first versions of our draft that helped us improve it. The work of F.~F. is funded by \textit{Estancias Posdoctorales por México, Estancia Posdoctoral Iniciales, Conahcyt-Secihti}. F.~F.~ acknowledges the support of DGAPA-PAPIIT UNAM, under Grant No. IN110622.  J.~M.~M.~ is indebted to Conahcyt-Secihti funding his Ph.D. P.~R.~ is partly funded by Conahcyt-Secihti (México), with the support of project CBF2023-2024-3226 being gratefully acknowledged, and by MCIN/AEI/10.13039/501100011033 (Spain), grants PID2020-114473GB-I00 and PID2023-146220NB-I00,
and by Generalitat Valenciana (Spain), grant PROMETEO/2021/071.
\label{sec:7}

\appendix
\section{Higgs decays $h\to \mu^{+} \mu^{-}+ \gamma/Z$}
\label{App:HDecay}
In this appendix we take advantage of the fact that, due to the large number of produced Higgses\footnote{A total of $\mathcal{O}(10^8)$ Higgs bosons for a $\sqrt{30}$ TeV Muon Collider with 90 $\text{ab}^{-1}$ of integrated luminosity.}, it is possible to study some rare decays, such as $h(p)\to \mu^{+}(p_1) \mu^{-}(p_2) \gamma/Z(p_3)$, which are produced by the same operators we previously analysed.

Taking into account the HEFT and SM tree-level amplitude, the contribution to the decay $h\to \mu^{+} \mu^{-} \gamma$ is exactly the same as the SMEFT case, since no other HEFT operator contributes, upon the global $a_\gamma$ parameter that could increase the total BR at most by a factor five. Considering $a_\gamma=5$, leads to a maximum value of ${\rm BR}(h\to \mu^{+} \mu^{-} \gamma)_{\rm NP}\approx 4\times10^{-9}$.

Meanwhile, for the decay $h\to \mu^{+} \mu^{-} Z$, we can study the effect that the NP under consideration could have, where all the HEFT operators contribute. In this case, the differential decay rate contributions are (invariant masses are defined in eq.~(\ref{eq_invmasses})): 
\begin{itemize}
\item Leading term in the SM: 
    \begin{align}
         \frac{d\Gamma}{dm_{13}^2 dm_{23}^2}=&\frac{g_Z^2 m_Z^2 (8\,{\rm sin}^4\theta_W-4\,{\rm sin}^2\theta_W+1)}{256\pi^3 m_h^3 v^2 (m_{13}^2 +m_{23}^2-m_h^2)^2}\Big( m_h^2 m_Z^2+2m_Z^4-2m_Z^2 (m_{13}^2+m_{23}^2)\nonumber\\&+m_{13}^2 m_{23}^2 \Big)\,.
         \label{SMleading}
    \end{align}
    \item $\text{Operator}\ \mathcal{N}_Z^{\mu}:$ 
    \begin{equation}
       \frac{d\Gamma}{dm_{13}^2 dm_{23}^2}= -\frac{C^2_{Z}}{\Lambda^2}\frac{(m_h^2 m_Z^2-m_Z^4+m_Z^2(m_{23}^2+m_{13}^2)-2 m_{13}^2 m_{23}^2)}{32\pi^3 m_h^3}\,.
    \end{equation}
     \item $\text{Operator}\ \mathcal{N}_2^{\mu}$ (this one does not bear a suppression factor $\propto 1/\Lambda$, making it the largest purely new physics effect, although much smaller than the SM-$\mathcal{N}_2^\mu$ interference): 
    \begin{equation}
        \frac{d\Gamma}{dm_{13}^2 dm_{23}^2}= \frac{C^2_{2}}{256 \pi^3 m_h^3 m_Z^2} \Big(m_h^2 m_Z^2+2m_Z^4-2m_Z^2(m_{13}^2+m_{23}^2)+m_{13}^2 m_{23}^2\Big)\,.
    \end{equation}
        \item $\text{Operator}\ \mathcal{N}_4^{\mu}:$ 
    \begin{equation}
         \frac{d\Gamma}{dm_{13}^2 dm_{23}^2}=\frac{C^2_{4}}{\Lambda^2}\frac{\big( m_h^2+m_Z^2-m_{13}^2- m_{23}^2\big)\big((m_{13}^2+m_{23}^2)^2-4m_h^2 m_Z^2\big)}{512\pi^3 m_h^3 m_Z^2}\,.
    \end{equation}
     \item $\text{Operator}\ \mathcal{N}_9^{\mu}:$ 
    \begin{align}
        \frac{d\Gamma}{dm_{13}^2 dm_{23}^2}=&\frac{C^2_{9}}{\Lambda^2}\frac{1}{512\pi^3 m_h^3 m_Z^2} \Big( 4 m_h^4 m_Z^2 + m_h^2 \big( -4 m_Z^4-4m_Z^2 (m_{13}^2+ m_{23}^2)\nonumber\\
        &+(m_{23}^2- m_{13}^2)^2\big)+m_Z^2(m_{13}^4+6 m_{13}^2 m_{23}^2+ m_{23}^4)-(m_{23}^2- m_{13}^2)^2\nonumber\\&(m_{13}^2+m_{23}^2)\Big)\,.
    \end{align}
    \item $\text{Interference}\ \mathcal{N}_Z^{\mu}-\text{SM}$ (numerically suppressed, both by the muon mass and because $4\, {\rm sin}^2\theta_W\sim1$): 
    \begin{align}
         \frac{d\Gamma}{dm_{13}^2 dm_{23}^2}=&-\frac{{\rm Re}(C_{Z})}{\Lambda} \frac{g_Z m_\mu}{64 \pi^3 m_h^3} \frac{(4\, {\rm sin}^2\theta_W-1)}{v\,m_{13}^2 m_{23}^2(m_{13}^2 +m_{23}^2-m_h^2)}\Big(m_h^4\big(m_Z^2(m_{13}^2+m_{23}^2)\nonumber\\&-m_{13}^2 m_{23}^2\big)+m_h^2\big(m_{13}^2 m_{23}^2(m_{13}^2+m_{23}^2)-m_Z^2(m_{13}^4+3 m_{13}^2 m_{23}^2+m_{23}^4)\big)\nonumber\\&+2 m_Z^2 m_{13}^2 m_{23}^2 \big(2(m_{13}^2+ m_{23}^2)-3m_Z^2\big)\Big)\,. 
    \end{align}
    \item $\text{Interference}\ \mathcal{N}_Z^{\mu}-\mathcal{N}_9^{\mu}$ (this is the only non-vanishing interference which is purely beyond the SM~\footnote{We do not consider purely beyond the SM interferences that are additionally suppressed by the muon mass.}, and thus suppressed): 
    \begin{align}
        \frac{d\Gamma}{dm_{13}^2 dm_{23}^2}= &-\frac{C_9 C_{Z}^*+C_{Z}C_9^*}{128\pi^3\Lambda^2 m_h^3} \Big( m_h^2 (4m_Z^2-m_{13}^2 -m_{23}^2)-m_Z^2 (m_{13}^2 +m_{23}^2)+(m_{23}^2-m_{13}^2)^2\Big)\,.
    \end{align}
     \item $\text{Interference}\ \mathcal{N}_2^{\mu}-\text{SM}$ (there is no parametric suppression $\propto 1/\Lambda$ in this term, which is potentially the largest new physics effect):
     \begin{align}
    \frac{d\Gamma}{dm_{13}^2 dm_{23}^2}= \frac{-C_2 g_Z {\rm sin}^2\theta_W}{64\pi^3 m_h^3 v (m_h^2-m_{13}^2-m_{23}^2)} (m_h^2 m_Z^2+2m_Z^4-2m_Z^2(m_{13}^2+m_{23}^2)+ m_{13}^2 m_{23}^2) \,.
\end{align}
        \item $\text{Interference}\ \mathcal{N}_4^{\mu}-\text{SM}$ (suppressed by the muon mass)
    \begin{align}
         \frac{d\Gamma}{dm_{13}^2 dm_{23}^2}&=\frac{- g_Z m_\mu\,{\rm Im}(C_4)}{256\pi^3\Lambda^2 m_h^3 m_Z^2 v m_{13}^2 m_{23}^2}\Big( m_h^4 m_Z^2 (m_{13}^2+m_{23}^2)-m_Z^2 m_{13}^2 m_{23}^2 (m_{13}^2+m_{23}^2) \nonumber\\
        &+m_h^2 \big\{ m_Z^4(m_{13}^2+m_{23}^2)-m_Z^2 (m_{23}^2-m_{13}^2)^2 - m_{13}^2 m_{23}^2 (m_{13}^2+m_{23}^2)\big\}\Big)\,.
    \end{align}
    \item $\text{Interference}\ \mathcal{N}_9^{\mu}-\text{SM}$ (again suppressed, because of $m_\mu$ and $4\, {\rm sin}^2\theta_W\sim1$): 
    \begin{align}
         \frac{d\Gamma}{dm_{13}^2 dm_{23}^2}&=\frac{g_Z m_\mu (4\, {\rm sin}^2\theta_W-1) {\rm Re}(C_9)}{256\pi^3\Lambda m_h^3 v m_{13}^2 m_{23}^2 (m_{13}^2+m_{23}^2-m_h^2)}\times \Big(m_h^6 (m_{13}^2+m_{23}^2)\nonumber\\
        &-m_h^4 (m_{13}^2+m_{23}^2) \big(m_Z^2+2 (m_{13}^2+m_{23}^2) \big)+m_h^2\Big\{m_Z^2(m_{13}^4-6 m_{13}^2 m_{23}^2+m_{23}^4)\nonumber\\
        &+(m_{13}^2+m_{23}^2)\big( m_{13}^4+4 m_{13}^2 m_{23}^2 +m_{23}^4 \big)\Big\}-3m_{13}^2 m_{23}^2 (m_{13}^2+m_{23}^2)\nonumber\\
        &(-2m_Z^2+ m_{13}^2+m_{23}^2) \Big)\,.
        \label{eqn:FinalInt}
    \end{align}
    \end{itemize}
We only took into account the leading order contributions, at most proportional to the lepton mass or quadratic in the HEFT coefficients. Also the substitution of the coefficients in eq. (\ref{eq:coefficients}) has to be done. The Mandelstam variables were defined as follows
\begin{equation}
    \begin{split}\label{eq_invmasses}
        &m_{13}^2\equiv(p_{\mu^{-}}+p_Z)^2=(p_h-p_{\mu^{+}})^2,\\
        &m_{23}^2\equiv(p_{\mu^{+}}+p_Z)^2=(p_h-p_{\mu^{-}})^2.
    \end{split}
\end{equation}
Once integrated, considering $\Lambda\approx 100$ TeV, the estimated values of every contribution to the BR$(h\to \mu^{+} \mu^{-} Z)$ are shown in table \ref{tab:BR}, where we first report the values without fixing the HEFT coefficients and then show the numerical results from taking  $|C_i^{\mu}|=10^{-4}$ and $\ a_i=1$, which is an optimistic approach where a large value of the Wilson coefficient is selected.
\begin{table}[h!]
\begin{center}
\begin{tabular}{|c|c|c|}
\hline
  Operator & $BR(h\to \mu^{+} \mu^{-} Z)$ &\makecell{$BR(h\to \mu^{+} \mu^{-} Z)$\\ $|C_i^{\mu}|=10^{-4},\ a_i=1$} \\ [5pt]
\hline
SM (Leading term) & $7.66\times10^{-4}$ & $7.66\times10^{-4}$ \\ [10pt]
\hline
$\mathcal{N}_2^{\mu}-$SM&$-2.52\times10^{-3}\ C_2^{\mu} a_2$  &$-2.52\times10^{-7}$  \\ [10pt]
\hline
$\mathcal{N}_Z^{\mu}-$SM&$-5.95\times 10^{-11}$ & $-5.95\times 10^{-11}$ \\ [10pt]
\hline
$ \mathcal{N}_2^{\mu}$& $5.29\times10^{-3}|C_2^{\mu}|^{2}a_2^2$ & $5.29\times10^{-11}$ \\ [10pt]
\hline
$\mathcal{N}_Z^{\mu} $& $9.02\times 10^{-13}$  & $9.02\times 10^{-13}$ \\ [10pt]
\hline
$\mathcal{N}_9^{\mu}-$SM& $-7.56\times 10^{-9}$ $\operatorname{Im}(C_9^{\mu})a_9$ & $-7.56\times 10^{-13}$ \\ [10pt]
\hline
$\mathcal{N}_4^{\mu}-$SM& $6.93\times10^{-9}\operatorname{Re}(C_4^{\mu})a_4$ & $6.93\times10^{-13}$ \\ [10pt]
\hline
$\mathcal{N}_Z^{\mu}-\mathcal{N}_9^{\mu}$& $2.08\times10^{-10}\operatorname{Im}(C_9^{\mu})a_9$ &$2.08\times10^{-14}$ \\ [10pt]
\hline
$\mathcal{N}_9^{\mu} $& $1.22\times10^{-8}|C_9^{\mu}|^{2}a_9^2$  & $1.22\times10^{-16}$ \\ [10pt]
\hline
$\mathcal{N}_4^{\mu} $& $2.15\times10^{-10}|C_4^{\mu}|^{2}a_4^2$ & $2.15\times10^{-18}$\\ [10pt]
\hline
\end{tabular}
\end{center}
\caption{\label{tab:BR} Estimated BR$(h\to \mu^{+} \mu^{-} Z)$ contributions within the HEFT approach.}
\end{table}

\setcounter{footnote}{0}
As can be seen from table \ref{tab:BR}, all the NP contributions are highly suppressed and many orders of magnitude below the SM leading term~\footnote{All of them are consistent with our previous explanations on their particular suppressions.}, even for the most optimistic HEFT-coupling cases. Thus, the main conclusion given in \cite{Buttazzo:2020ibd} for the SMEFT case still holds here, being that the current muon $g$-2 anomaly cannot be tested at a muon collider through this decay process~\footnote{The interference $\mathcal{N}_2^{\mu}-$SM term could be accessible only if high statistics were achieved in a future muon-collider and all relevant radiative corrections for the SM contribution to that order were known.}.  

These results were somewhat expected, since the main advantage of the NP scattering process is that it grows with energy and thus can become dominant over the SM cross-section at a very high-energy collider, in contrast with the decay process, fixed at the Higgs mass scale. 

Then, the HEFT contribution in the $h\to \mu^{+} \mu^{-} +\gamma/Z$ channel is still too small to be observed, being the scattering process the most promising scenario to test the $g$-2 NP contribution in a model-independent way, highlighting the potential of a muon collider in high-energy physics.

\bibliographystyle{unsrt}
\bibliography{biblio.bib}
\end{document}